\documentclass[preprint]{aastex}

\begin{document}

\title{Can gravitational waves be detected in quasar microlensing?}

\author{Shane L. Larson}
\affil{Department of Physics, Montana State University, Bozeman, Montana
59717}

\author{Rudolph Schild}
\affil{Harvard-Smithsonian Center for Astrophysics, 60 Garden Street, 
Cambridge, MA  02138}

%\recieved{}
%\accepted{}

\begin{abstract}
Studies of the lensed quasar ${\rm Q}0957+561 {\rm A,B}$ have shown
evidence for microlensing in the brightness history of the quasar
images.  It had been suggested that a frequency offset between the
brightness fluctuations in each of the two images might possibly be
caused by gravitational radiation generated by a massive black hole
binary at the center of the lensing galaxy.  This paper demonstrates
that the fluctuations produced by such a source of gravitational waves
will be too small to account for the observed frequency offsets.
\end{abstract}


\section{Introduction}\label{sec:Intro}

The ${\rm Q}0957+561 {\rm A,B}$ system was the first recognized
multiple image gravitational lens (Walsh, Carswell, and Weymann 1979)
and the first to have a measured time delay (Schild and Cholfin 1986). 
Once the time delay between the two images was measured, the
brightness histories of the two images could be compared and ${\rm
Q}0957$ was the first to show evidence for a microlensing event
(Vanderriest et al.\ 1989; Schild \& Smith 1991; Pelt et al.\ 1998). 
The long brightness records accumulated to measure the time delay and
microlensing at optical (Schild and Thomson, 1997 and earlier
references therein) and radio (Haarsma et al., 1997) wavelengths have
been analyzed by Thomson and Schild (1997) to look for other artifacts
in the time series data.

A number of such artifacts were found, including evidence for
multipaths, coherency in the radio and optical, weak sinusoidal
behavior on timescales of weeks, and a puzzling frequency offset
between the coherent brightness fluctuation in the A and B images. 
Since the A and B images are of the same quasar, lack of frequency
coherence of the two images shifted by the measured time delay
suggests that some process related to the microlensing has probably
altered the frequency.

The possibility that the frequency offset might be caused by binary
microlensing objects was illustrated in Schild and Thomson (1993).  A
more intriguing possibility has also been suggested (Schild and
Thomson 1997): the frequency offset might constitute the detection of
gravitational waves radiated by a binary black hole at the nucleus of
the lens galaxy.  The purpose of this paper is to explore this
possibility more carefully.

The possibility that a massive binary black hole might reside at the
center of the lens galaxy is reinforced by what is known about the
monster.  It is a very luminous elliptical galaxy near the center of a
cluster, suggesting it is probably a starpile type galaxy, resulting
from a long history of mergers of smaller galaxies.  The lens galaxy
is a known radio source (Greenfield et al.\ 1980) so it is presumed to
contain at least one black hole at the center.  If the galaxy is a
starpile type, then it would have accumulated one or more massive black
holes from galactic mergers.  Dynamical friction with the constituent
stars in the merged galaxy would cause the massive black holes to
settle into the central density cusp over the course of time.  If one
were to look for binary black holes, this is the place one would start
the search.  The nearest analogue to the lens galaxy would be M87,
near the center of a subgroup of the Virgo cluster.  Evidence for
massive black hole binaries has been found in other astrophysical
systems; for example a binary black hole has been inferred to exist
in quasar Q1928+738 (Roos, Kaastra, and Hummel 1993).

To picture the physical configuration for a gravitational wave
explanation consider a terrestrial observer looking in the direction
of a distant quasar.  The observer sees two images because the
gravitational field of a lens galaxy near the line of sight creates
two light paths to the observer.  As the quasar light passes through
the lens galaxy, stars within the galaxy act as microlenses that can
further magnify the luminous quasar structure.  Any dark matter
objects would also introduce microlensing.  The theory for the
formation of two quasar images indicates that the surface mass density
is sufficiently high that microlensing events should always be
underway.  If a gravitational wave originating at the center of the
lens galaxy passes through the field of microlenses, it should alter
the pattern of null geodesics by which light passes through the lens
galaxy, and impose the wave's periodicity on the resulting signal,
suggesting the gravitational wave's frequency should be evident in the
time series record of the quasar's brightness.

This paper is organized as follows.  Section 2 summarizes the evidence
for the observed Q0957 frequency offset.  In Section 3, the amount
that microlensing signals would be affected by propagating
gravitational waves is calculated, and we conclude that no observable
signature would be found.  Section 4 reconsiders the origin of the
observed effect in Q0957 and speculates on prospects for observing
gravitational waves in other microlensed quasar systems.

Geometric units where $G = c = 1$ are used throughout; conventional
units are adopted (and specified) when numerical values are quoted.

\section{Frequency Shifts in the Q0957 Coherence}\label{sub:FreqShift}

Frequency shift effects in ${\rm Q}0957$ were first discussed in
Thomson and Schild (1997), who noted that (p.\ 188) ``Analyzing the A
and B light curves jointly, we find a frequency shift between them. 
This shift is one-sided, so it is not the result of a simple
modulation, and has about the same value as the periodicity seen in
the microlensing.''  A more comprehensive discussion and a plot of the
unexpected behavior is given in Schild and Thomson (1997), who plot
the dual-frequency cross coherence between the delayed A and the B
light curves for a $3286-$day time interval.  Their surprising result
is that although the two images originate in the same quasar, the
brightness records have almost no cross-coherence over the frequency
band $1.5 - 5.0$ cy/yr.  Instead, the coherency seems to be shifted up
to higher frequencies, commonly 2 to 3 cy/yr higher than the expected
frequency.  Because the quasar's random brightness fluctuations appear
to be phase coherently shifted to higher frequency over a restricted
frequency band, the effect might be thought of as a kind of blueshift
that operates over only a limited frequency band.  Several such blue
shifting frequencies are found, and it is not yet clear whether a
single such blueshift operates over only a limited time interval (a
limited range of Julian dates).  In this report we focus on the
dominant frequency offset, 0.29 cycles/year.  Because the affected
frequency band was shown by Schild (1996) to be dominated by
microlensing effects, it seems likely that the shift of frequency
coherence results from the microlensing.

For modulation of the microlensing signal by gravitational waves to be
observable, two conditions must be met.  First, the amplitude of the
gravitational wave at the microlensing field must be large enough to
alter the microlensing amplitude, implying the amplitude must be a
significant fraction of the Einstein radius of a microlensing
particle.  Second, the quasar source must also have sufficiently fine
structure that the background to the wave-altered geodesics produces
the periodically modulated brightness changes.

While the quasar accretion disc is usually taken to be quite large and
structureless, a study of the Q0957 statistics by Schild (1996) showed
that fluctuations were seen to be so rapid that they evidenced dark
matter in the form of condensed objects (rogue planets) having
approximately terrestrial mass and cosmologically significant numbers. 
They also evidenced compact quasar structure, which seemed surprising
at the time but has since been confirmed in the observation of an
apparent cusp crossing in quasar ${\rm Q}2237$.\footnote{See the data
at website
http://www.astro.princeton.edu/$\sim$ogle/ogle2/huchra.html} These
Q0957 and Q2237 observations place important constraints on the size
of the emitting quasar source.  For the brightness profile illustrated
in Schild's (1996) Fig.\ 5, a time scale of $60$ days is evidenced,
about the same as the peak of the power spectrum illustrated in his
Fig.\ 4.  Thus quasar structure on size scales less than the accretion
disc diameter seems to be indicated by the microlensing data for two
quasars.

Because the B quasar image passes 5 times closer to the lens galaxy
nucleus than A, we presume that gravitational waves would be greatest
in B, and compute the spacing of the microlensing masses to determine
the amplitude of gravitational wave that would be needed.  The
normalized surface mass density for the B image was defined in Schmidt
and Wambsganss (1998) and shown to be equal to $1.17$.  Schmidt and
Wambsganss also show that for standard cosmology, the angular size
diameter of a solar mass Einstein ring projected to the quasar source
is $3 \times 10^{13}$ cm.  The calculated $1.17$ surface mass density
means that along any line of sight, on average an observer sees one
microlensing mass magnifying the background quasar source, so on
average the Einstein rings are adjacent to one another (they touch
each other with $17\%$ overlap, loosely speaking).  Because the
distance to the lens is nearly half the distance to the quasar in a
correct calculation of the angular diameter distances, the separation
of the microlenses projected onto the plane of the sky at the distance
of the lens galaxy is about half the Einstein ring diameter, or about
$10^{13}$ cm.

From these numbers we wish to estimate an amplitude scale for which
the gravitational waves emanating from the putative black hole at the
center of the lens galaxy would periodically influence the
microelnsing.  We presume that the gravitational waves passing the
network of microlenses alter the null geodesics of the quasar beams
periodically, and cause the observed periodic changes in the
microlensing magnification.  The microlensing magnification is usually
illustrated as the pattern of high magnification cusps that the masses
impose upon the light beam from the distant source.  Such cusp
patterns have been illustrated, for example, by Wambsganss (1992) and
Seitz, Wambsganss, and Schneider (1994), who show that for optical
depths near 0.1 the cusps have a characteristic size scale about the
same as the Einstein ring diameter, but for higher optical depths,
such as $1.17$ appropriate for Q0957, the cusp scale is about a factor
of ten smaller, especially when the random motions of the field
microlenses are taken into account (Wambsganss \& Kundic, 1995).  Thus
we take the characteristic size scale for the brightness amplification
cusps to be $10^{12}$ cm.  We are aware that this is also aproximately
the size of the putative black hole at the center of the quasar.

Because the brightness fluctuations observed are small, only 3\% or
so, it may not be justified to assume that the gravitational wave
induced alteration of the quasar light's path is the full amplitude of
the cusp pattern; even an alteration of the propagation path by 10\%
of the characteristic cusp spacing should produce an effect of the
small amplitude observed.  If the predicted amplitude of the
propagation path alteration were $10^{11}$ cm, we would feel obliged
to model this and sharpen this estimate, but we shall see that the
predicted effect falls short of this amplitude by several orders of
magnitude.  In this sense, we take the fiducial amplitude for any
gravitational wave alteration of the pattern of null geodesics through
the pattern of microlensing masses to be $10^{11}$ cm.

\section{The expected gravitational wave amplitude}\label{sec:Amplitude}

To explicitly compute the effect of a gravitational wave on photons 
propagating through a symmetric Schwarzschild gravitational lens, 
consider the geometry shown in Figure \ref{fig.Geometry}.  An 
Earth-bound observer and a distant source of photons are situated 
about a thin gravitational lens, with the source lying at a 
misalignment angle $\beta$ from the line of sight to the lens.  The 
Einstein radius is given by 
\begin{equation}
   \eta = \sqrt{{2 M} \over {L}}\ ,
   \label{EinsteinRadius}
\end{equation}
where $M$ is the lens mass and $L$ is the separation between the lens 
and observer (or lens and source).

If one wants to consider the effect of the wave on photons propagating
through the lens and toward a distant observer, one approach is to
consider the superposition of a linearized Schwarzschild metric
(describing the lens) and a linearized metric describing gravitational
plane-waves.  Consider the case of a gravitational wave with amplitude
$h$, frequency $\omega$ and $+$ polarization propagating through a
Schwarzschild lens down the $+x-$axis.  The spacetime metric may be
written as
\begin{equation}
   ds^{2} = \left(1 - {{2 M} \over r}\right) dt^{2} - \left(1 + {{2
   M} \over r}\right) \left(dx^{2} + dy^{2} + dz^{2}\right) + h \cos
   \omega(t - x) \left(dy^{2} - dz^{2} \right)\ ,
   \label{Metric}
\end{equation}
where $r$ is the radial coordinate from the lens.  Note that in the 
limit $h \rightarrow 0$, this becomes the Schwarzschild metric in 
isotropic coordinates.

A standard (but computationally intensive) approach to studying photon
trajectories in the spacetime described by the metric of Eq.\
(\ref{Metric}) is to write out and solve the null geodesic 
equations.  Such a study would allow one to ascertain what, if any, 
observable effect the gravitational wave might have on an observed 
microlensing signal.

A simpler approach may be used in the thin lens approximation, where 
the photon trajectories may be approximated as straight lines 
deflected when they pass through the lens plane.  In this case, an 
application of Fermat's principle can be used to extremize the time 
of flight through the lens in lieu of solving the geodesic equations.  
The condition that the time of flight be an extremum will provide the 
necessary geometrical information to consider how the microlensing 
signal is affected by the gravitational wave.

Fermat's principle in non-stationary spacetimes has been discussed in 
the specific context of gravitational lensing (Kovner 1990; 
Nityananda and Samuel 1992), and has been used to examine the effect 
of cosmological gravitational waves on the time delay between lensed 
quasar images (Frieman, Harari and Surpi 1994, hereafter FHS).  The 
conclusion of FHS was that for long wavelength gravitational waves, 
any time delay between the images due to the gravitational wave would 
be interpreted by observers as part of the misalignment angle 
$\beta$.  We will follow the example of FHS, using Fermat's principle 
to compute the apparent misalignment angle $\beta(h)$ as a function 
of the gravitational wave amplitude.

FHS showed that for the lens/gravitational wave system described by
the metric in Eq.\ (\ref{Metric}) the time of flight may be written
\begin{equation}
    \tau \simeq \int_{-L}^{+L} dz \left[1 + {1 \over 2} \left({{dx} 
    \over {dz}}\right)^{2} + {1 \over 2}h \cos \omega(t - x) + {{2 
    M} \over r} \right]\ ,
	\label{TimeFlight}
\end{equation}
where $(dx/dz)$ characterizes the photon paths in Fig.\ 
\ref{fig.Geometry}.  

Extremizing the time of flight in Eq.\ (\ref{TimeFlight}) leads to an 
expression for the deflection angle $\alpha$ in the lens plane:
\begin{equation}
    \alpha = 2 \left(\theta - \beta_{g} \right)\ ,
	\label{Deflection}
\end{equation}
where $\theta$ is the apparent location of the image, and $\beta_{g}$ 
is the apparent misalignment angle in the presence of the 
gravitational waves, which was found by FHS to be
\begin{equation}
   \beta_{g} = -{ h \over {\omega L}} \sin^{2}\left( {{\omega L} 
   \over 2} \right) \sin \omega \left( t_{e} + L \right)\ .
   \label{betag}
\end{equation}
This expression for $\beta_{g}$ is valid so long as $\eta L \ll 
1/\omega$ ({\it i.e.}, as long as the wavelength of the gravitational 
wave is much larger than the typical size of the Einstein radius of 
the microlens).

In general, increasing the misalignment angle reduces the lensing
effect.  In microlensing, the lens mass is small enough that
individual images are unresolvable, and increasing the misalignment
angle reduces the microlens amplification.  In the presence of
gravitational waves, the apparent misalignment angle is given by Eq.\
(\ref{betag}).  The condition that $\eta L \ll 1/\omega$ amounts to
requiring that the gravitational wave vary on a timescale much less
than the time of flight for photons through the vicinity of the lens. 
This suggests that over a long period of time, the microlensing
amplification may change due to the fact that $\beta_{g}$ is a slowly
varying function of time.

Any change in $\beta_{g}$ over the course of time amounts to a change
in the impact parameter of a given photon trajectory which passes
through the lens.  The misalignment angle subtends an arclength
\begin{equation}
    s = L \beta_{g} =  -{ h \over {\omega}} \sin^{2}\left( {{\omega L} 
   \over 2} \right) \sin \omega \left( t_{e} + L \right)
	\label{arclength}
\end{equation}
for a lens which lies a distance $L$ from the observer.  Eq.\
(\ref{arclength}) can be separated into a scaling amplitude (given by
$h/\omega$) multiplied by a periodic function which varies between
$-1$ and $+1$.  The quantity of interest is the maximum value of $s$,
which we take to be
\begin{equation}
   s_{max} \sim {h \over \omega}\ .
   \label{maxS}
\end{equation}
$s_{max}$ represents the maximum deflection of a null geodesic
(passing through a thin gravitational lens) by a gravitational wave of
amplitude $h$ and frequency $\omega$.

The gravitational wave amplitude can be estimated by assuming the case
of an equal mass circularized binary source.  At a distance $r$ from a system
with total mass $2m$, and gravitational wave frequency $\omega$, the 
amplitude is well approximated by
\begin{equation}
    h \sim {{(2 m)^{5/3} \omega^{2/3}} \over r}\ .
	\label{GWAmplitude}
\end{equation}
Combining this with Eq.\ (\ref{maxS}) yields
\begin{equation}
   s_{max} \sim {{(2 m)^{5/3}} \over {r \omega^{1/3}}}\ .
   \label{maxS2}
\end{equation}

Eq.\ (\ref{maxS2}) may be evaluated for the case of Q0957.  Assuming
the anomalous frequency shift in the brightness history is caused by
gravitational waves, the $3.4$ yr period implies a gravitational wave
frequency\footnote{This value for $\omega$ is the frequency of the
emitted gravitational radiation (given by the observed $3.4$ yr
period), corrected for the redshift of the source galaxy, $z = 0.37$.}
of $\omega \sim 8 \times 10^{-8}$ Hz.  The separation between the
microlens and the source of gravitational waves is taken to be the
linear distance between the quasar image which is being microlensed
and the center of the primary lensing galaxy.  For the Q0957 system,
the distance between the B image and the nucleus of the lens galaxy
has been measured to be $1.045$ arcsec (or $r \sim 5$ kpc for a
lensing galaxy at a redshift of $z \sim 0.37$, assuming a flat
cosmology with a Hubble parameter of $75$ km s$^{-1}$ Mpc$^{-1}$)
(Bonometti, 1985).  If the typical mass of a central galactic black
hole is $m \sim 10^{8} M_{\odot}$, then the maximum deflection is
\begin{equation}
    s_{max} \sim 10^{6} {\rm cm}\ .
	\label{sValue}
\end{equation}

\section{Discussion}\label{sec:Discussion}

This calculation of the maximum deflection of the pattern of null
geodesics due to a gravitational wave from a putative binary black
hole at the center of the lens galaxy shows that no observable effects
would be expected.  The gravitational wave deflects the propagation
paths by $10^{6}$ cm, whereas a deflection of order $10^{11}$ cm (the
smallest size scale expected to be found in quasar source structure)
would be required to explain the frequency offset between the A and B 
images.

Might the effects be observable in a different lens system having
different parameters?  Our Eq.\ (\ref{maxS2}) shows that the
deflection amplitude is a weak function of the gravitational wave
source frequency $\omega$, which is presumed known for the Q0957
system; other lens systems with a factor 100 higher binary black hole
frequency are certainly possible (they would have a lifetime shorter
than a Hubble time).  The Q0957 system has the widest image separation
of all known lens systems, so most other lenses would have a smaller
value of $r$, the impact parameter of the quasar beam passing the
binary black hole; thus an effect might be slightly favored in other
gravitational lens systems (Eq.\ (\ref{maxS2})).  An effect would be
strongly favored by a more massive black hole binary, but we have
already used large values for the orbiting masses.  We conclude that
the detection of gravitational waves with quasar microlensing seems
not to be possible for any known system.

We are left to specualte about the nature of the periodicity observed
in the microlensing (Schild and Thomson 1997).  A microlensing
explanation seems essential because the observed effect is
asymmetrical; the frequency offset seems to affect the B image most
strongly, and the B image has the hugest optical depth to
microlensing.  With the gravitational wave explanation excluded it
seems most likely that the source of the periodic activity is orbital
effects in the microlensing objects themselves.  This would literally
mean that the planetary mass objects found by Schild (1996) to
dominate the mass of the lens galaxy are in some (or most) cases
binaries.  Heretofore we have disfavored this explanation because a
simple calculation of the transverse velocities predicts that the
microlens' orbital velocity would be much less than such cosmological
velocity components as the motion of the sun around the center of the
galaxy (220 km/sec), or the motion of the source quasar or lens galaxy
relative to its local Great Attractor (600 km/sec).  Nevertheless it
would be possible for some orbital effects to be introduced into the
microlensing, and more intricate microlensing simulations might be
necessary to explore an orbital explanation.

\acknowledgments The authors would like to thank David J.\
Thomson, Chris Kochanek and William A.\ Hiscock for helpful
discussions.  The work of SLL was supported by NASA Cooperative
Agreement No.\ NCC5-410.

\newpage

\begin{figure}
  \caption{The geometry of image formation for the QSO's B image, past
  a hypothetical microlens situated 10 kpc from the center of the G1
  lens galaxy.  The microlensing produces a doubling of the quasar's B
  image with an image separation of $10^{-6}$ arcsec, so the images
  are not resolved; however the passage of a gravitational wave causes
  periodic fluctuations in the deflection angle $\beta_{g}$ which
  might cause variations in the lens magnification and thus impose a
  periodic frequency offset in the observed pattern of quasar
  brightness fluctuations.}
  \label{fig.Geometry} 
\end{figure}


\begin{thebibliography}{9}

\bibitem[Frieman, Harari \& Surpi 1994]{FHS94} Frieman, J., Harari,
D.\ and Surpi, G., 1994, \prd, 50, 4895

\bibitem[Greenfield, Roberts \& Burke 1980]{GRB80} Greenfield, P.,
Roberts, D., and Burke, B. 1980, Science 208, 495

\bibitem[Haarsma et al.\ 1997]{H97} Haarsma, D. et al., 1997, \apj,
479, 102

\bibitem[Kovner 1990]{kov90} Kovner, I., 1990, \apj, 351, 114

\bibitem[Nityananda \& Samuel 1992]{nit92} Nityananda, R.\ and Samuel, 
J., 1992, \prd, 45, 3862

\bibitem[Pelt et al. 1998]{PSRS98} Pelt, J.,
Schild, R., Refsdal, S., and Stabell, R., 1998, \aap, 336, 829

\bibitem[Roos, Kaastra \& Hummel 1993]{RKH93} Roos, N., Kaastra, J.,
and Hummel, C., 1993, \apj, 409, 130

\bibitem[Schild 1996]{S96} Schild, R., 1996, \apj, 464, 125

\bibitem[Schild \& Cholfin 1986]{SC86} Schild, R.\ and Cholfin, B.,
1986, \apj, 300, 209

\bibitem[Schild \& Smith]{SS91} Schild, R.\ and Smith, R.\ C., 1991
\aj, 101, 813

\bibitem[Schild \& Thomson 1993]{ST93} Schild, R.\ and Thomson, D.\
J., 1993, ``Gravitational Lenses in the Universe: Proceedings of the
31st Liege International Astrophysical Colloquium'', ed.\ J.\ Surdej et
al., p.\ 415

\bibitem[Schild \& Thomson 1997]{ST97} Schild, R.\ and Thomson, D.\ J.,
1997, ``Astronomical Time Series'', ed.\  D.\ Maoz, A.\ Sternberg, and 
M.\ Leibowitz [Kluwer: Dordrecht], p.\ 73

\bibitem[Schmidt \& Wambsganss 1999]{SW99}Schmidt, R.\ and Wambsganss,
J., 1999, \aap, 335, 379

\bibitem[Seitz, Wambsganss, & Wchneider 1994] {SWS94} Seitz, C. Wambsganss,
J. and Schneider, P., 1994, \aap, 288, 19

\bibitem[Thomson \& Schild 1997]{TS97} Thomson, D. J., and Schild, R.
1997, ``Applications of Time Series in Astronomy and Meteorology'',
ed.\ T.\ Subba Rao, [Chapman and Hall: New York], p.\ 187

\bibitem[Vanderriest et al.  1989]{V89} Vanderriest, C.\ et al., 1989,
\aap 215, 1

\bibitem[Walsh et al.\ 1979]{WCW79} Walsh, D.\ ,
Carswell, R.\ and Weymann, R., 1979, \nat, 279, 381

\bibitem[Wambsganss 1992]{W92} Wambsganss, J. 1992, \apj, 392, 424

\bibitem[Wambsganss \& Kundic 1995]{WK95} Wambsganss, J. and Kundic, T.,
1995, \apj, 450, 19


\end{thebibliography}
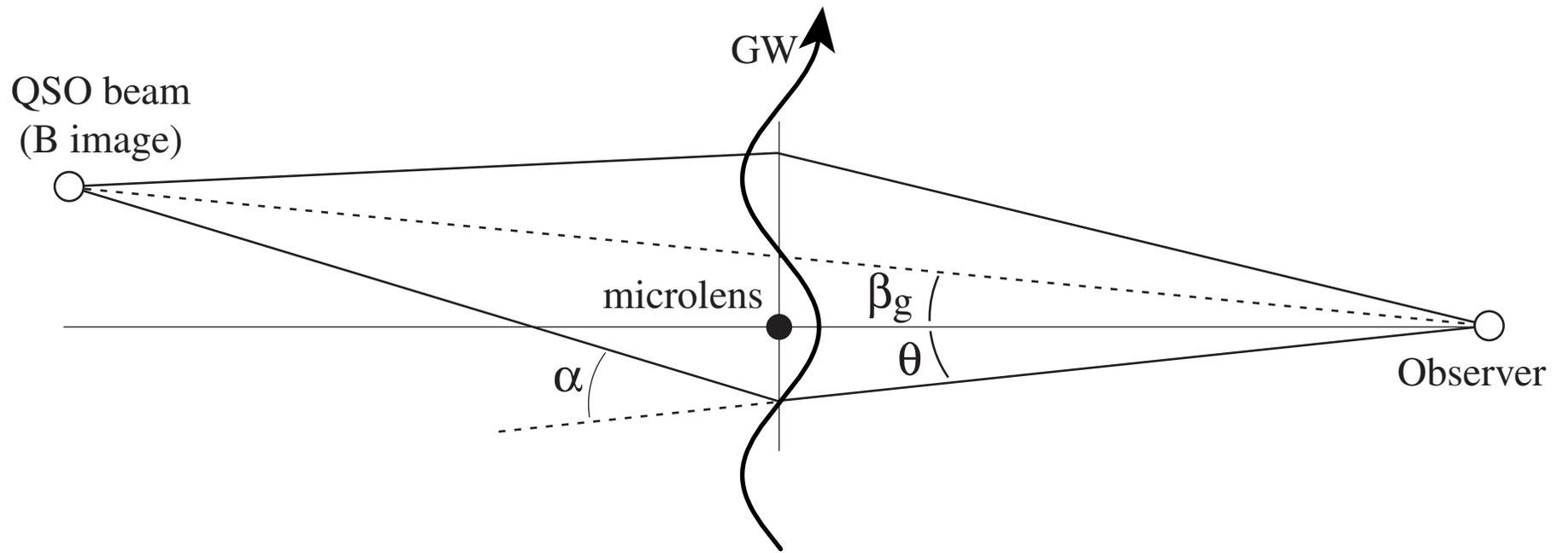
\end{document}